\documentclass[useAMS,usenatbib,usegraphicx]{mn2e}

\title[Self-Regulated Winds]{The Self-Regulated Winds of Long
Period Variable Stars}

\author[C. Struck et al.] {Curtis
Struck$^{1}$\thanks{E-mail: curt@iastate (CS);
lwillson@iastate.edu (LAW)}, Daniel C. Smith$^{2}$,
Lee Anne Willson$^{1}$\footnotemark[1], 
\newauthor
Gary Turner$^{3}$, and George H. Bowen$^{1}$\\ 
$^{1}$Dept. of Physics and Astronomy, Iowa State University, Ames, IA
50014 USA\\
$^{2}$L-3 Communications Analytics Corp., 1655 Forest Hill Ct.,
Crofton, MD 21114\\
$^{3}$Dept. of Mathematical Sciences, Morningside College, Souix
City, IA 51106}

\begin{document}


\pagerange{\pageref{firstpage}--\pageref{lastpage}} \pubyear{2003}

\maketitle

\label{firstpage}

\begin{abstract}

Numerical models of the dynamically extended atmospheres of long
period variable or Mira stars have shown that their winds have a very
simple, power law structure when averaged over the pulsation cycle.
This structure is stable and robust despite the pulsational wave
disturbances, and appears to be strongly self-regulated.
Observational studies support these conclusions.  The numerical models
also show that dust-free winds are nearly adiabatic, with little
heating or cooling.  However, the classical, steady, adiabatic wind
solution to the hydrodynamic equations fails to account for an
extensive region of nearly constant outflow velocity.  An important
process or group of processes is missing from this solution.  Since
gas parcels moving out in the wind are periodically overrun by
pulsational waves, we investigate analytic solutions which include the
effects of wave pressure, heating, and the resulting entropy changes.

In the case of dust-free winds we find that only a modest amount of
wave pressure is needed to derive an analytic model for a steady,
constant velocity, locally adiabatic outflow.  Wave pressure is
represented with a term like that in the Reynolds turbulence equation
for the mean velocity.  The waves damp relatively quickly with radius,
as a result of the work they do in driving the mean flow.  Although
the pressure from individual waves is modest, the waves are likely the
primary agent of the self-regulation of the dust-free winds.

In dusty Miras, the numerical models show the radiation pressure on
grains and the subsequent momentum transfer to the gas, play the
dominant roles in driving the wind, and wave pressure is not very
important.  In the models of the dusty wind, the gas variables also
adopt a power law dependence on radius. Heating is required at all
radii to maintain this flow, and grain heating and heat transfer to
the gas are significant.  Both hydrodynamic and gas/grain thermal
feedbacks can transform the flow towards particular self-regulated
forms.

\end{abstract}

\section{Introduction}

The astrophysical theory of hydrodynamic winds began with the solar
wind equations proposed by \citet{par58}, also see
\citep{par63}. Parker's model predicted that the pressure gradient
between the sun's outer atmosphere and the interstellar medium would
drive the wind.  In subsequent years Parker's solutions to the
hydrodynamic equations have proven to be a powerful tool for studying
stellar winds in general. (The analogous \citet{bon52} solutions for
spherical accretion problems have proven to be equally important.)

\citet{bir64a, bir64b, bir64c} constructed shock heated models of the
solar corona and wind, which can be regarded as Parker winds with an
additional heating term.  However, Bird's models were not designed to
capture the highly nonlinear shock dynamics of long period variable
star atmospheres (henceforth LPVs).  The large-amplitude,
quasi-ballistic motions behind LPV shocks were not considered (see
\citealt{wil79} and \citealt{hil79}).  Nonetheless, a number of
important results relevant to Miras were anticipated, including:
acoustic heating of the circumstellar gas, a dynamic balance between
shock heating and expansion cooling in the wind, and self-regulation
in the wind flow. Bird proposed these as driving forces for the solar
wind and corona, which are now believed to be the result of
magnetohydrodynamic dissipation (e.g., the review of
\citealt{par97}). At the present time, these processes appear more
relevant to the warm extended atmospheres and winds of dust-free LPVs
(see the reviews of \citealt{wil97}, \citealt{wil00} for discussions
of these structures).  As will be described below, self-regulating
processes are very important, but Bird's conjecture that the
self-regulation works to maintain a constant Mach number throughout
the flow is not strictly correct.

Hartmann and MacGregor (\citealt{har80}, \citealt{har82}, also see
\citealt{mac97}) proposed an application of solar corona and wind
models with Alfv\'{e}n wave heating and pressure to LPV winds. These
processes are similar to the shock pressure and heating processes
described below. The Hartmann and MacGregor models assumed an
approximately hydrostatic corona, with waves viewed as a modest
perturbation, in contrast to the strong shocks observed in
LPVs. Specifically, these are polytropic wind models, without shock
entropy generation. Both shock and Alfv\'{e}n waves may play a role in
LPV winds, but we believe that the strong pulsational shocks are
dominant, so we will not consider magnetic effects in this paper.

Because LPVs have very large amplitude pulsations, they might be
expected to provide difficulties for smooth wind solutions of the
hydrodynamic equations.  Indeed, the numerical models of \citet{bow88,
bow90}, and \citet{bow91} show that the atmospheres of pulsating Miras
are highly dynamic. This is also true of all published numerical
models of these and related stars, e.g., \citet{fle92}, \citet{feu93},
\citet{hof97}, \citet{ste98}, \citet{win00}, and \citet{hof03}. Most
of these models do not include non-LTE radiative cooling, so in the
remainder of this paper we will primarily refer to Bowen's models, and
the particular runs described below.  See \citet{wil97} for a
comparison of various models.)

In the numerical models, the density near the photosphere retains a
roughly exponential decline, but departs from this to a power law
dependence at density value that depends on the mass outflow rate. In
the transition from exponential to power law decline, the dynamical
lifting by shocks followed by (in some cases) radiation pressure on
grains lifts and accelerates the material. In this flatter density
profile, more gas has been lifted to large radii, and this along with
radiation pressure on dust grains in some cases, provide the means for
driving much enhanced winds.  Bowen's models show that the inner
atmosphere is a complex region where shock acceleration and heating
and cooling all play important roles (see \citealt{wil97}).  On the
other hand, the wind region at large radii has a simpler structure,
and the gas variables averaged over pulsation period can be well
approximated as simple power law functions of radius (see Figures 1-3
and discussion below).  Such simple profile forms suggest that it
should be possible to construct simple analytic models in the
tradition of Parker and Bird, at least for the smooth wind region.

\begin{figure*}
\scalebox{0.4}{ \includegraphics{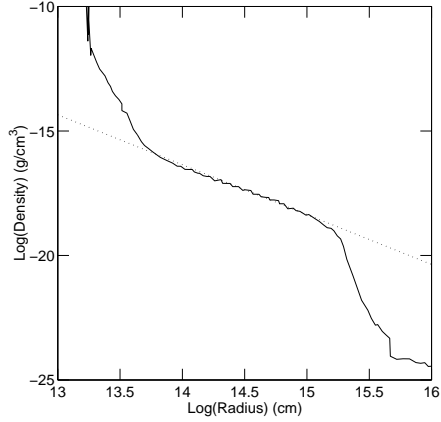}}
 \caption{Density profile from a dust-free, hydrodynamical model like
those of \citet{bow88}. The model was run for a very long time to
allow a large steady wind region to develop and relax out to $r \simeq
10^{15} cm$. (See text for details). The dotted line shows an
a $\rho \propto r^{-2}$ function (normalized to the numerical model at
log(r) = 14.5) for comparison.}
 \label{f1}
\end{figure*}

\begin{figure*}
\scalebox{0.4}{ \includegraphics{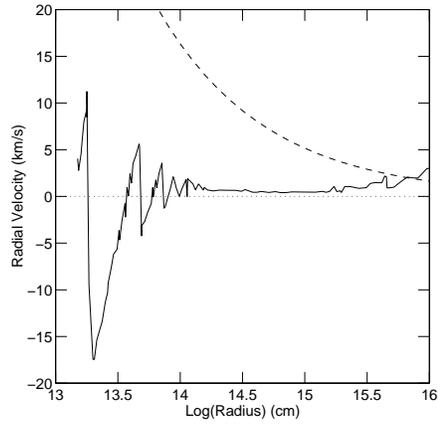}}
 \caption{Radial velocity profile from the end of the same dust-free
Bowen model shown in the previous figure. Comparison to the dotted
line at zero velocity shows the low, constant value of the outflow in
the steady wind region. The upper dashed line shows the local escape
velocity. The lower dashed curve shows a fit derived from an analytic
model,  see Sec. 2.6 for details.} 
 \label{f2}
\end{figure*}

\begin{figure*}
\scalebox{0.4}{ \includegraphics{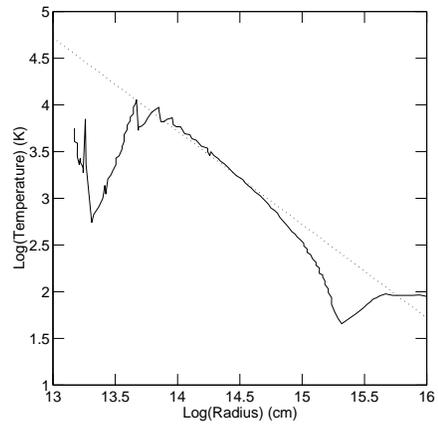}}
 \caption{Temperature profile from the end of the same long, dust-free
Bowen model shown in the previous figures. The dotted line shows an a
$T \propto r^{-1}$ function (normalized to the numerical model at
log(r) = 14.5) for comparison.}
 \label{f3}
\end{figure*}

We will not consider LPV observations in detail in this paper, since
our main goal is to understand the regularities revealed by numerical
models.  In addition, the constraints on theory provided by
observation (e.g., spectra) are indirect, and generally the comparison
between theory and observation is best done with the aid of detailed
numerical models (see e.g., \citealt{wil00}, \citealt{tej03} and
references therein). Nonetheless, some recent observations provide
quite direct information about the gas variables in the winds and
extended atmospheres of LPVs, and a brief mention of these provides a
context for the subsequent work.

First of all, a major assumption of both analytic and numerical models
is that the LPV winds are approximately spherically symmetric. In
spite of evidence for modest deviations from symmetry, or asymmetries
in binary systems, there is much evidence from a variety of wavebands
for approximate symmetry in most cases (e.g., \citealt{buj91}).

Recent infrared interferometric observations of ``dust shells'' around
Miras are also generally consistent with spherically symmetric models,
though not with uniform dust density distributions.  The data are
better fit by models including a few distinct shells where the dust
emission is high (see e.g., \citealt{hal97}, \citealt{lop97},
\citealt{mon97}, \citealt{fon03}, and references therein).  However,
these observations probe a region located at a distance of only a few
stellar radii, where conditions are described as dynamic and complex.
In fact, the numerical models show that this is the region where
shocks have grown to very nonlinear amplitudes, where dust forms if it
is able to, and where the wind is just beginning to be accelerated
(the ``shock acceleration'' and ``dissipation'' zones of
\citealt{wil97}). Thus, these observations generally support the
picture provided by the models, and do not contradict the notion of a
generally smooth wind structure.

Observations and models are in general agreement about basic wind
parameters like the mass loss rate or flow velocity. From CO
observations (e.g., \citealt{buj91}, \citealt{kah94}, \citealt{you95},
\citealt{ker98}, \citealt{gro99}, and \citealt{win02}, also see
\citealt{ala01} for mid-infrared results) we know that mass loss rates
range from $10^{-7} M_{\odot}/yr$ up to $10^{-5} M_{\odot}/yr$ for
Miras in agreement with Bowen's models (\citealt{bow88},
\citealt{bow91}, \& \citealt{wil00}). Above about $10^{-5}
M_{\odot}/yr$ the stars are generally classified as OH-IR sources. In
this paper we are most interested in low mass loss rates, in
relatively dust-free cases.

The dust free models develop an approximately constant outflow
velocity with a speed well below the escape speed, and this pattern
persists over many stellar radii. In this region the flow is also
subsonic. Thus, this constant velocity is not the same as the coasting
flow in standard wind models well beyond the sonic point. The
observations also find low wind velocities (of order 5-10 km/s)
and little evidence for changes in the flow velocity through the
wind. For the Miras CO observations are able to probe the flow at
large distances from the star, and so, provide important constraints
on any model. Interestingly, Cepheid wind velocities also seem to
be low relative to the escape speed (\citealt{sas94a,sas94b}).

In the following sections we will derive analytic wind models,
described by hydrodynamic equations which including approximate,
averaged terms, like those in the Reynolds turbulence equations, to
represent the effects of nonlinear pulsational waves on the flow. We
then discuss the self-regulatory feedbacks in these winds, and use
several approaches to argue that these processes select constant
velocity outflows over other possible wind solutions.

\section{Wave Pressure in Dust Free Winds}
\subsection{Classical Steady Wind Solutions Compared to a Numerical
Model}

The usual Parker wind equations are derived from the steady,
spherically symmetric hydrodynamic continuity and momentum equations,
which can be written as,
\begin {equation}
\dot{M} = 4 \pi r^{2} \rho u
\end {equation}
\begin {equation}
u\frac{du}{dr}=-\frac{1}{\rho}\frac{dP}{dr}-\frac{GM}{r^{2}}.
\end {equation}
where the variables, $\dot{M}$, $M$, $\rho$, $P$, $u$, and $r$ denote
the mass loss rate, total mass of the star, the density, pressure,
wind speed, and the distance from the center of the star,
respectively.  Classical steady wind solutions also assume a
polytropic equation of state, $P = K\rho^{\gamma}$, with constant K,
$\gamma$.  In this paper we will assume a constant rate of mass loss
as well.

Two properties of the outer region of the atmosphere suggest that it
can be described as approximately locally adiabatic: 1) the
temperatures there are in a range where radiative cooling is
inefficient, and 2) the shocks in this region are relatively weak.
The latter feature is evident in Figures 1-3, which show density,
velocity, and temperature profiles taken from the end of a very long
run (about 1000 pulsation cycles) of Bowen's modeling code, without
dust.

Detailed descriptions of the atmospheric dynamics can be found in
\citet{bow88}. Here we merely note that the code numerically solves
the one-dimensional, spherically symmetric, hydrodynamic equations
with a \citet{ric67} type Lagrangian method, including a conventional
artificial viscosity algorithm for stabilizing strong shocks. A number
of different chemical rate, momentum and energy sources are calculated
explicitly for each gas element.  These include: the atomic radiative
cooling rates, the free electron abundance, energy changes due to
ionization and recombination. They also include simple approximations
to the radiative transfer, grain formation, radiation pressure on
grains, and grain heating and cooling. The effects of most of these
processes will not be considered in this paper, which focusses instead
on understanding the global hydrodynamics. However, the numerical
models provide a standard for comparison, as well as assurance that
the analytic models are not unrealistic. The parameters of the
numerical model shown in Figures 1-3 are essentially the same as those
of a dust-free, $1.0 M_{\odot}$, ``standard'' model of \citet{bow88},
though the model shown here was computed with increased spatial
resolution. The numerical models will be compared to analytic models
below, especially in Section 2.6.

The wind density profile shown in Figure 1 has an approximately
$1/r^2$ form, while the form of the corresponding temperature profile
is nearly $1/r$ (Fig. 3). Note that beyond a radius of $10^{15} cm$ in
the figures the model wind flow is not fully relaxed. (Indeed, Fig. 3
suggests that the flow is not thermally relaxed beyond $log(r) =
14.5$.) There is, in fact, a polytropic solution to equations (1) and
(2), with a $1/r^2$ density profile, and a $1/r$ temperature
profile. However, the value of the polytropic index in this model is
$\gamma = 3/2$. Not only does this not correspond to the adiabatic
value of $\gamma = 5/3$ expected in the absence of cooling and
heating, it does not have any obvious physical meaning. 

We should note, however, that Parker argued against values of $\gamma
> 3/2$ on the grounds that larger values did not yield ``a solution
beginning at low velocity close to the sun and extending outward to
zero pressures at infinity'' (\citealt{par63}, pg. 61). He also noted
the need for a heating source when $\gamma < 3/2$.

What about the adiabatic, $\gamma = 5/3$ wind solution?  To illustrate
the problems with this classical solution we integrate equation (2)
from an inner radius $r_{1}$ to an arbitrary outer radius r, use the
ideal gas law and employ the definition of the adiabatic sound speed
to derive the following Bernoulli equation,
\begin {equation}
\frac{1}{2}
\biggl(u^{2}-{u_1}^{2}\biggr) = \frac{-1}{\gamma - 1}
\biggl(c^{2} -{c_1}^{2}\biggr) +
\frac{1}{2} \biggl({v_e}^{2} - {v_{e1}}^{2}\biggr).
\end {equation}
The numerical models show that the sound speed c decreases with radius
approximately as $r^{-1/2}$, as does the escape velocity
$v_{e}$. Thus, according to the equation (3), the flow velocity u must
decrease with radius by comparable amounts.  It does not; Bowen's
models show constant velocity in the wind at all times (see Fig. 2).

We note also that, if all the wind flow lies on a single adiabat in
the thermodynamic phase space, and has a $1/r^2$ density profile, then
the ideal gas law implies that the temperature goes as $T \propto
r^{-4/3}$. With the long run we have performed, which yields a large
wind region, the numerical profile does appear shallower than an
adiabatic one.

Maintaining a shallower temperature profile requires extra heating and
momentum sources in the wind.  In fact, a likely reason for
discrepancies between numerical models and the adiabatic solution is
that, although the pulsational shocks are weak in the wind region,
their momentum, heat, and entropy inputs are not negligible.

\subsection{Not Quite Adiabatic Solutions with Wave Pressure and Heating}
\subsubsection{Wave Pressure}

The question is how to capture the effects of waves in a relatively
simple analytic description of the overall wind flow, averaged over
many pulsation cycles?  The problem is like that of describing
turbulent flows in which large fluctuations on many scales coexist
with a mean flow and long-lived coherent structures.  The
characteristics of the regular, pulsational shocks, which propagate
outward through the wind, are very different from the stochastic,
fluctuations described in the theory of well-developed, homogeneous
turbulence (e.g., \citealt{ten72}, \citealt{mcc90}).  Nonetheless, in
both cases we are dealing with disturbances on small spatial scales
and short timescales, which we don't necessarily need to resolve in
order to describe their average effect on the flow.

In fact, the fundamental idea of turbulence theory, that the flow can
be divided into a mean and a fluctuating part, provides a very
appropriate approach to generalizing the classical stellar wind
equations.  We cannot simply adopt the averaging procedures used to
derive the Reynolds equations of turbulence theory, since these depend
on the random nature of the fluctuations in homogeneous turbulence.
We can, however, provide physical justifications for including similar
'averaged' fluctuation terms in a set of phenomenological Reynolds
equations for mean wind flows.

The numerical models show that each pulsational shock provides a
compression and an outward boost to an individual gas element, and the
gas element follows a quasi-ballistic free-fall trajectory behind the
shock (\citealt{wil79}, \citealt{hil79}, \citealt{bow88}, and the
review of \citealt{wil97}). More precisely, for a periodic pattern of
motion in the atmosphere, the material at larger r must have smaller
outward postshock speeds. This picture is good deep in the
atmosphere. If the postshock speed is slightly too high, the mass
element encounters the next shock at slightly larger than it did the
last one. This gives a small net outward motion, as is seen in the
atmosphere around 1.5-2.5 stellar radii. Where the flow dominates, the
material still responds nearly ballistically, setting up a stable
pattern relative to the mean flow speed and providing a small
increment of momentum to the gas in each cycle. This allows us to
separate the mean flow from the oscillations, which average
approximately to zero (see Appendix A for details).

To incorporate the effects of the shocks on the mean flow, an
additional term is included in the (inviscid) momentum equation. With
the approximation of the previous paragraph, this term can be written
as the divergence of a Reynolds stress, which in the spherically
symmetric case is just the radial derivative of the mean square
velocity fluctuation, or ${\partial}(\sigma^2)/{\partial}r$.
Physically, this is the (one-dimensional) wave pressure gradient. In
the Bernoulli equation (3) it contributes a $\sigma^2$ term, like the
sound speed term.

\subsubsection{Shock Heating}

The net effect of shock heating can be included as a term in the mean,
steady state energy equation (see Appendix A), which can be written
as,

\begin {eqnarray}
\frac{1}{2}r^{-2}\frac{\partial}{\partial r}
\biggl(r^{2} \rho u^{3}\biggr)
& = & - r^{-2}\frac{\partial}{\partial r}
\biggl(r^{2} \rho u \epsilon\biggr) +
r^{-2}\frac{\partial}{\partial r}
\biggl(r^{2} u P\biggr)
\nonumber\\
&& - \frac{GM \rho u}{r{2}} + \Gamma\rho.
\end {eqnarray}
The variables, $\epsilon$ and $\Gamma$ denote the internal energy per
unit gas mass, and the shock heating function, respectively.
Henceforth, we define u as the sum of the mean ($U = <u>$) and
fluctuating ($\sigma = (u-U)$) velocity components. In turbulence
theory, we assume that this is an ensemble average over many nearly
identical systems, distinguished by the details of their random
fluctuations. In the present case the average is also assumed to be
over all pulsational phases. Similarly, $\rho, \epsilon,$ and P
generally have both mean and fluctuating parts, but we will not need
to adopt special symbols for the separate parts (except in Appendix
A).

We re-emphasize that in this equation we neglect the effects of the
interaction between wind material and the stellar radiation, the
interaction with the radiation generated in different parts of the
winds, and energy exchanges between gas and dust grains.

The Reynolds formalism provides an additional equation for the mean
square fluctuating velocity (see McComb, sec. 1.3.2).  In the
case of a steady, inviscid, constant mean velocity radial flow, this
equation can be written,
\begin {eqnarray}
U \frac{\partial}{\partial r}
\biggl(\sigma^{2}\biggr)& = &
-\frac{\partial}{\partial r} \Biggl[<(u-U)^{3}> + \frac{2}{\rho}
<(u-U) P>\Biggr] 
\nonumber\\
&& -2{\sigma}^2 \frac{\partial U}{\partial r}
 - 2\Gamma.
\end {eqnarray}
The first two terms on the right-hand-side represent turbulent (or
wave) energy diffusion by nonlinear couplings, which we expect to fall
off quickly with radius in this case, and so will neglect them. (This
neglect is part of a closure approximation for the Reynolds moment
equation set.) What remains is a relation between the local wave
pressure gradient and the shock heating,
\begin {equation}
\Gamma = - \frac{1}{2} U \frac{\partial}{\partial r}
\Bigl(\sigma^{2}\Bigr) + {\sigma^{2}} 
\frac{\partial U}{\partial r},
\end {equation}
(where the last term is small in winds with nearly constant mean
velocity). 

Although we now have the required energy and momentum source terms and
a relation between them, the equation set is not quite complete.

\subsubsection{Equation of State}

We employ a locally, but not globally, adiabatic equation of state to
describe the effects of shock heating in the outer atmosphere.
Physically, in each pulsation cycle, any parcel of gas goes through an
irreversible (non-closed) cycle in a p-V (pressure - specific volume)
phase diagram.  A pulsational shock wave pushes a gas parcel along a
Hugoniot curve in the phase diagram, and off its initial adiabat
(e.g.,\citealt{lan59}).  Radiative cooling at constant pressure behind
the shock then moves it to lower specific volume (i.e., compresses
it).  Downstream from the shock the parcel evolves along a new adiabat
to higher specific volumes and lower pressures in the rarefaction
region, until it is hit by the next shock. A possible laboratory
analogue of this physics is provided by the reverberating shock
cavities produced in the (gun-type) experiments (see \citealt{hol95}).

For weak shocks the specific entropy changes are not great, but they
accumulate. Gas parcels farther out in the wind have experienced more
of these cycles, and so, lie on adiabats that are farther from the
initial one than those in the inner wind. We assume that the specific
entropy of the mean flow is time-independent and a smooth function of
radius, like the other gas variables. We believe this gradual entropy
change is an important factor in making the wind region of Bowen's
dust-free Mira models appear quite adiabatic, but with a constant
outflow velocity, which is not characteristic of a classical adiabatic
wind.

We can write the locally adiabatic equation of state as,

\begin {equation}
P = K(r) \rho^{\gamma},
\end {equation}
where $\gamma = 5/3$. Because gas parcels at different radii are on
different adiabats, K is a function of the radius, rather than a
constant. (\citet{kla03} have also recently considered model Keplerian
disks with entropy gradients due to local heating.)

Since the gas is locally adiabatic, the perfect gas relationship for
$\epsilon$ is valid (locally) at any radius,
\begin {equation}
\epsilon = \frac{P}{\rho} \frac{1}{\gamma - 1}.
\end {equation}

This relation completes the equation set. We have introduced three new
terms into the Parker wind equations, which describe wave pressure,
shock heating, and shock entropy production. None of these extra
phenomenological terms are found in the gas equations that are the
basis of published numerical models (e.g., \citealt{bow88}). Those
models follow the time-dependent pulsation and shock phenomena
explicitly. We believe all of the phenomenological terms are necessary
to describe the locally adiabatic {\it mean} flow in pulsationally
driven winds.

We note that the classical, linear treatment of waves propagating into
a stellar atmosphere indicates that long period waves should be
reflected at the surface. Pijpers and collaborators, described the
effects of stochastic acoustic waves in detail in a series of papers
(\citealt{pij89a}, \citealt{pij89b}, \citealt{kon93}, \citealt{pij93},
and \citealt{pij95}). They found that such waves can propagate in the
context of a general outflow (see \citealt{pij93}). \citet{bow90}
found that the power needed to maintain large amplitude photospheric
oscillations is actually lower for long periods, and these induce
nonlinear wave propagation (shocks) traveling out through the
atmosphere. Here we take the existence of transmitted waves as a
given, and focus on the large-scale, nonadiabatic effects of mildly
nonlinear waves in the wind region only.

\subsection{Constant Velocity Winds and Entropy Production}

Motivated by the observations, we begin by considering the simple case
in which the mean wind velocity is constant, $U = U_0$. In this case,
equation (1) immediately gives $\rho$, and the mean flow momentum
(eq. (2)) can be written,

\begin{equation}
-\frac{1}{\rho}\frac{dP}{dr}-\frac{GM}{r^{2}}
-\frac{d}{dr}(\sigma^{2})
= 0,
\end {equation}
where $\rho$ and P are now the Reynolds average quantities.

With the additional assumption that all three terms on the
left-hand-side of this equation scale the same way with radius
throughout the wind flow, we can readily obtain a solution to this and
the energy equation. This assumption appears quite strong, but is
physically reasonable for the following reason. The wind extends over
a very large range of radii. The gas thermal energy and temperature do
not fall off as rapidly with radius in dust-free models as expected in
an adiabatic flow. This suggests a heat source operates throughout;
shock heating is the only available source. (We note that
photo-heating of grains will have a similar scaling with radius, but
we assuming negligible grain populations for the present.) This fact
ties the thermal pressure to the shocks and suggests that it scales
like the wave pressure.

Beyond this physical argument, we can briefly note the mathematical
possibilities. First is the possibility that thermal pressure
dominates in part of the wind and wave pressure elsewhere. In the
former region we will have a Parker thermal wind, with the
observational difficulties mentioned above. In the wave dominated
region, we will get the same scaling with radius, which derives from
balancing the gravitational term. Since shock compression does provide
thermal heating, this limit isn't physically self-consistent. A second
possibility is that the thermal and wave pressures have different
functional forms whose sum contrives to balance the gravitational
term. For example, they could both have the same power law form
derived below, but modulated by oscillatory parts which have the same
amplitude for both, but which are $180^{\circ}$ out of phase.  Any
other solution of this type would have to be similarly
fine-tuned. Given that the wind extends over orders of magnitude in
radial extent, this seems physically contrived. We will argue later
that the solution below is also preferred by self-regulatory
processes.

Specifically, we assume that $\sigma^{2} = A/r$ for some constant A,
then equation (9) becomes,

\begin{displaymath}
dP = - \Bigl(G M - A\Bigr)
\frac{\rho}{r^{2}} dr.
\end{displaymath}

Now $\rho$ can be substituted from the continuity equation and the
result can be integrated (from a radius r to $\infty$) to get

\begin{equation}
P = \frac{1}{3} \frac{(GM-A)}{r} \rho,
\end{equation}
assuming $\rho$ and P go to zero at infinite radius. It is interesting
that this equation has the form $P \propto \rho^{3/2}$ of a constant
velocity Parker wind, since $\frac{1}{r} \propto \rho^{\frac{1}{2}}$,
though we here view the solution as only locally adiabatic, rather
than globally barotropic.

Next, we want to simplify the energy equation (4) in this constant
velocity case. We need the following result in taking the mean of the
term on the left-hand-side of equation (4),
\begin{equation}
\Bigl<\rho{u^3}\Bigr> = \Bigl<\rho\Bigr>
\Bigl<{U^3} + 3{\sigma^2}U\Bigr>
\end{equation}
where $\rho$ is used here as the total density (mean plus fluctuating
parts). This equation assumes that the mean of odd powers of the
fluctuating velocity vanish, and that the correlation $<\rho(u-U)>$ is
negligible (see Appendix A). Henceforth, we drop the brackets $<>$,
and assume again that $\rho$ and P refer to the mean quantities.

When we use this result, pull $U_0$ factors out of partial
derivatives, and substitute for $\epsilon$ from equation (8), the
energy equation becomes,
\begin{eqnarray}
\frac{3}{2}\Biggl(\frac{\rho U_{0}}{r^{2}}\Biggr)
\frac{\partial}{\partial r}
\Bigl({r^2}{\sigma^2}\Bigr) +
\frac{3}{2}\Bigl({U_0}{\sigma^2}\Bigr)
\frac{\partial {\rho}}{\partial r}& = &
\nonumber\\
\Biggl(\frac{\gamma-{2}}{\gamma-{1}}\Biggr) \frac{U_0}{r^2}
\frac{\partial {\Bigl({r^2}P}\Bigr)}{\partial r} -
\frac{GM{\rho}{U_0}}{r^2} +
\Gamma \rho,&&
\end{eqnarray}

To further simplify we can substitute for P with equation (10), for
$\Lambda$ with equation (6), use the substitution ${\sigma}^2 =
A/r$. The result reduces to,
\begin{displaymath}
A = {\frac{5}{11}}GM,
\end{displaymath}
and yields in turn,
\begin{equation}
\Gamma = \frac{5}{22} U_0 \frac{GM}{r^2}.
\end{equation}

The physical interpretation of the last equation is that the shock
heating rate is doing work against the gravitational potential $GM/r$
at each radius, on a flow timescale of about $r/U_0$. It is reducing
the effective gravity (i.e., the factor GM-A in eq. (10)), that must
be overcome by flow down the thermal pressure gradient.

Combining equation (10) with equation (7) and the continuity equation
gives the following expression for the variable $K$ in equation (7).

\begin{equation}
K(r) = \frac{2}{11} G M
\Biggl(\frac{\dot{M}}{4 \pi U_0}\Biggr)^{-\frac{2}{3}}
r^{\frac{1}{3}},
\end{equation}

\noindent and that equation becomes,

\begin{equation}
P =  \frac{2}{11} G M
\Biggl(\frac{\dot{M}}{4 \pi U_0}\Biggr)^{-\frac{2}{3}}
r^{\frac{1}{3}} \rho^\gamma,
\end{equation}
a generalized (position-dependent) polytropic relation.

To better appreciate why K varies with radius, recall that the entropy
(per unit mass) of a perfect gas is,

\begin{equation}
S = log\Bigl({P/\rho^\gamma}\Bigr),
\end{equation}
so equation (7) implies $S = \log(K)$. Thus, K is directly related to
entropy, and its radial dependence implies a mean radial entropy
gradient. {\it{As any gas element flows outward, it is overtaken by
shocks moving through the wind, and each shock increases the entropy
of the element.}} At any fixed radius the entropy production is
balanced by the outward transport.

The value for the constant A above implies that the wave energy is
about 45\% of the gravitational potential energy at all radii.  This
implies that $\sigma$ should be about half the value of the escape
velocity at any radius. The velocity ''jitter'' due to shocks in
Bowen's dust-free numerical models is considerably less than this,
generally less than a third of the escape velocity.

However, it is not clear that this velocity jitter should be directly
identified with $\sigma$, see section 2.6 below. We expect
difficulties in capturing the full range of velocity variation in the
models. Most of this variation occurs across thin shocks, and these
short wavelength variations are not well resolved numerically.

\subsection{Accelerating Winds}

The constant velocity wind described above is an especially simple
solution to the equations.  In this section we consider steady
accelerating winds with radially dependent mean velocities. We
restrict our consideration to power law similarity
solutions. Mathematically, we are essentially exploring a specific
kind of variation of the previous solution, but self-similar winds are
the most relevant physically.

Specifically, we assume $U = U_0 (r/r_0)^\delta$, so $\rho =
\frac{\dot{M}r_0^\delta}{4 \pi u_0} r^{-(2+\delta)}$.  In this case,
the mean momentum equation is,

\begin{equation}
U\frac{dU}{dr} =
-\frac{1}{\rho}\frac{dP}{dr}-\frac{GM}{r^{2}}
-\frac{d}{dr}\Bigl(\sigma^{2}\Bigr).
\end {equation}

To solve this equation we will again assume a specific form for the
wave pressure term, that is,

\begin{equation}
\sigma^{2} = \frac{A}{r} + B
\Biggl(\frac{r}{r_0}\Biggr)^{2\delta},
\end{equation}
where the first term is as in the constant velocity case, and the
second term is designed to adjust the wave pressure in accordance with
the flow acceleration.  Like A, the coefficient B is independent of
radius, but does depend on the exponent $\delta$.  Note that in the
case $\delta = 0$, this expression does not coincide with the previous
one for constant flow unless we add an additional term of -B.  This
term complicates the algebra of the model considerably, and since we
usually have $|B| << A/r_0$ (see below), we will neglect it for the
moment.

We could also add other power law terms to the expression for the wave
pressure, or pursue a series solution for the wind variables. We
believe this would only complicate the equations without adding much
physical insight.

An expression for the pressure can now be derived by integrating the
momentum equation as before. We obtain,

\begin{eqnarray}
P(r)& = & \Biggl(\frac{1}{3 + \delta}\Biggr)
\Bigl(GM-A\Bigr) \frac{\rho(r)}{r} +
\nonumber\\
&&\Biggl(\frac{\delta}{\delta-2}\Biggr)
\Biggl(1 + \frac{2B}{U_0^2}\Biggr) {U_0^2} \rho(r)
\Biggl(\frac{r}{r_0}\Biggr)^{2\delta}.
\end{eqnarray}

\noindent The function K(r) can be derived by substituting equation
(7) for P.  

Next we substitute equation (18) (with nonconstant U) into the
${\sigma}^2$ equation (5) (including the term for the flow of energy
from the mean field to the fluctuating velocity field, $-2{\sigma}^2
\frac{\partial U}{\partial r}$). This produces a rather complex
expression for the heating,

\begin{equation}
\Gamma = \frac{U_0}{2r_0}
\Biggl[\Bigl(1 - 2\delta\Bigr)
\frac{A}{r_0}
\Biggl(\frac{r}{r_0}\Biggr)^{\delta - 2} -
4\delta B
\Biggl(\frac{r}{r_0}\Biggr)^{3\delta - 1}\Biggr].
\end{equation}

Finally, the above expressions can be substituted into the energy
equation to derive the following expressions for the coefficients A
and B as functions of the exponents $\gamma$ and $\delta$ by equating
the coefficients of the equal powers of $\delta$,

\begin{equation}
A = \frac{\Bigl[(2-\gamma) - (3+\delta)(\gamma-1)\Bigr]}
{\Bigl[(2-\gamma) - (2-\delta)(3+\delta)(\gamma-1)\Bigr]} GM,
\end{equation}

\begin{equation}
B = \frac{\Bigl[-(\gamma-1)(2-\delta) + (2-\gamma)2\delta\Bigr]}
{\Bigr[5(\gamma-1)(2-\delta) - 4\delta(2-\gamma)\Bigr]} {U_0^2}.
\end{equation}

The equations (20)-(22) specify the net heating required to maintain
the power law velocity profile. They do not directly take into account
the dependence of specific heating processes, like shock heating, on
the gas quantities, but rather assume that such processes can be
regulated to the form above. However, the expression for the heating
profile (eq. (20)) is so complex that it seems unlikely that realistic
shock heating and radiative cooling processes would generate it. This
is in contrast to the simple heating profile of equation (13) for the
constant velocity outflow, and suggests that the simpler form would be
preferred (as it is in Bowen's numerical models). This question will
be examined more carefully in the next section.

Before continuing, however, we note a couple of peculiarities of the
equations above. First of all, in the limit of $|\delta| << 1$ and
$\gamma = 5/3$ (locally), the expressions for A and B become,

\begin{equation}
A = \frac{5}{11} GM \biggl(1 + \frac{32}{55}\delta\biggr),
\end{equation}

\begin{equation}
B = -\frac{1}{5} {U_0^2}
\biggl(1 - \frac{3}{10}\delta\biggr).
\end{equation}

\noindent B is negative, and not directly proportional to
$\delta$. This means that in the accelerating wind, the wave pressure
could be reduced at $r < r_0$ relative to the constant velocity flow.
Physically, it is more reasonable to simply identify $r_0$ as the
inner radius of the wind (see eq. (18)).

Another peculiarity is that if, as we expect, the heating is dominated
by the first term in equation (20), it becomes negative when $\delta$
is increased above a value of 1/2, which implies a cooling process.
This also seems unphysical.

\subsection{Self-Regulation}

\subsubsection{General Considerations}

The simple appearance of the profiles of gas variables in Bowen's
dust-free numerical models suggests that the system relaxes to a
self-similar solution, independent of the details of the shock
heating, and subsequent expansion cooling. We have suggested above
that one key to deriving this solution is the assumption that the gas
in the wind is {\it only locally adiabatic}. This assumption allows
for entropy production in weak shocks, which gradually evolves a gas
element through a sequence of adiabats as it moves outward in the
wind.

The thermal state of the gas determines the environment traversed by
the shocks, which in turn are responsible for producing this state.
This feedback in the shock heating mechanism eventually works to
regulate the wind to the $1/r^2$ density profile.

The process of self-regulation is well illustrated by the early
relaxation that occurs in Bowen's numerical models, which are started
with small (but increasing) pulsation amplitudes, and with an
exponential atmospheric density profile. As described in
\citealt{bow88} the traveling wave parts of the pulsations steepen
into strong shocks as they propagate down the initial exponential
atmospheric profile. These shocks launch gas parcels out on
quasi-ballistic trajectories (\citealt{wil79} and \citealt{hil79}) to
such large radii that they are not able to fall back to their starting
points before being hit by the next shock. As a result, the initial
profile is stretched out into power law form. This causes a feedback,
the shock acceleration is decreased in the flatter pressure profile,
and the ballistic launch velocities are moderated. Thus, if the
pressure profile is steeper than $1/r^3$, then the shock amplitude
grows, pushing material out, until it matches that form (see Appendix
B). If the density gradient is shallower, however, the loss of energy
in pushing the gas will lead to declining shock amplitudes, and it
will not be possible to hold the shallow profile up against gravity. 
A pressure gradient that is neither too steep, nor too flat, also
maintains the flat outflow velocity.

The steady gas variable profiles that are eventually established with
finite amplitude pulsations include another region between the
exponential atmosphere where shocks grow nonlinearly, and the
power law wind region. This is the shock dissipation region where the
power law density profile flattens, but the temperature and pressure
increase with radius to a maximum, forming a warm ``Calorisphere'' and
the inner boundary of the wind. See the review of \citet{wil97} and
references therein for more details. The development of this region in
the numerical models provides a clear example of the effects of shock
dissipation when density and pressure profiles flatten.

This qualitative discussion on relaxation processes, which was
partially anticipated by \citealt{bir64c}, helps us understand why
there is a preferred wind profile. However, it is not yet sufficiently
precise to predict the form of that profile. Next we will consider
some more quantitative approaches to the problem.

\subsubsection{Nonequilibrium Thermodynamics and Least Dissipation}

The relaxation to the simple, constant velocity flow is driven, in
part, by the tendency of such a system towards a state of least
dissipation, or minimum entropy production. Such states often
correspond to those of maximum entropy content. The Theorem of Minimum
Entropy Production for steady, nonequilibrium states derives from work
of Helmholtz, though it has been generalized and used in a variety of
applications in the last few decades (see e.g., \citealt{gla71},
\citealt{pri80}, \citealt{woo96}). Proofs of the different versions of
this theorem involve substantial restrictions, e.g., that the system
is not too far from thermodynamic equilibrium, or that there are
linear relations between the flow variables and their driving force or
source terms (like the familiar linear stress-strain relation), see
\citet{woo96}.

These theorems have not been used much in astrophysics, where there
are many non-equilibrium systems, but where these systems are often
highly time-dependent. On the other hand, the winds of long-period
variables appear to provide an astrophysical situation where the
theorem can be usefully applied. If pulsations begin with small
amplitudes and build up steadily, then the wind can adjust to this
changing driving, and never find itself too far from a set of local
equilibria. Moreover, equation (9) (neglecting the last term) appears
to give the linear relation required of the dissipative quantities in
some proofs (see \citealt{woo96}). The minimization of shock
dissipation also seems to be in accord with the qualitative
considerations of the previous subsection.

We can apply the theorem by comparing the net production of the
entropy in the accelerating wind models. The theorem implies that the
wind with minimal entropy production is the preferred state. In
principle, we need to consider the details of the dissipational
processes, which is very complex in the present case. Fortunately, it
seems clear physically that the entropy production in the steady winds
must be directly related to the shock heating function $\rho\Gamma$
(see eq. (20)). Thus, for the purposes of making a simple estimate, we
define a function, H, as follows,

\begin {equation}
H = \Biggl(\frac{2}{\dot{M}}\Biggr)
\int_{r_0}^{r_1}
4\pi {r^2} \rho\Gamma dr
\end {equation}

\noindent where $r_0$ and $r_1$ are adopted inner and outer bounds of
the wind region. The coefficient $(2/{\dot{M}})$ is included merely to
absorb that common factor from the continuity equation into the
definition of H, which then has units of velocity squared.
Substitution for $\rho\Gamma$ and integration yields,

\begin {equation}
H = \Bigl(1-2\delta\Bigr)
\frac{A(\delta)}{r_0}
\Biggl(1 - \frac{1}{x_1}\Biggr)  +
2B(\delta)
\Bigl(1-{x_1^{2\delta}}\Bigr),
\end {equation}

\noindent where the functions A and B are given by equations (21) and
(22) with $\gamma = 5/3$, and $x_1 = {r_1}/{r_0}$. Note that the outer
boundary of the wind, ${r_1}$, will generally be a function of
$\delta$, which governs the density falloff. For example, we can make
a simple estimate by assuming that the outer boundary is where the
mean flow velocity U equals the escape velocity. In that case, we
have,

\begin{equation}
x_1 = \frac{r_1}{r_0} =
\Biggl(\frac{2GM}{{r_0}{U_0^2}}\Biggr)^{\frac{1}{1+2\delta}}.
\end{equation}

The dimensionless factor $({U_0^2}{r_0})/GM$, which appears both in
the ratio of coefficients in equation (26) $B{r_0}/A$ and in equation
(27), is the primary similarity parameter of the problem. For
convenience, we name it,

\begin{equation}
Z_0 = \frac{{U_0^2}{r_0}}{GM}.
\end{equation}

Now, using equations (21), (22), and (27) we can approximate H as,

\begin{eqnarray}
H& = & \frac{GM}{r_0}
\frac{(1-2\delta)(5+2\delta)}
{\bigl(11+2\delta(\delta -1)\bigr)} \Biggl(1 -
\biggl(\frac{Z_0}{2}\biggr)^{\frac{1}{1+2\delta}}\Biggr) +
\nonumber\\
&&2{U_0^2}
\frac{(-2+2\delta)}{(10-7\delta)}
\Biggl(1 - \biggl(2Z_0\biggr)^
{\frac{2\delta}{1+2\delta}}\Biggr).
\end{eqnarray}

Then we take the derivative with respect to $\delta$, and set it equal
to zero to find the dissipation extrema. The resulting expression is
very complicated, and not worth recording here without some
simplifying manipulations and approximations.

As yet, the only extra approximation we have used in deriving the
expression for H (or its derivative) is the outer boundary estimate of
equation (27). However, at this point, it is helpful to adopt the
approximation that ${U_0^2} << \frac{GM}{r_0}$, i.e., that the
inner-edge wind velocity is much less than the inner-edge escape
velocity. This assumption is clearly satisfied in the numerical
models. Since $B \propto {U_0^2}$, and $\frac{A}{r_0} \propto
\frac{GM}{r_0}$, this is equivalent to the assumption that $|B| <<
A/{r_0}$ already noted above. However, it is not wise to neglect
the B-term in equation (26) above if $\delta$ is small and negative,
because the term ${x_1^{2\delta}}$ can be relatively large.

Next, consider limits on the magnitude of $\delta$. When $\delta$ is
positive, the first term in equation (29) is larger than the second,
and for $\delta > 1/2$ it is negative. Then the dissipation H would
be negative, which is unphysical. Thus, we do not need to consider
large, positive values of $\delta$. Similarly, when $\frac{-5}{2} <
\delta < \frac{-1}{2}$ the first term makes H negative, and
unphysical. For large negative values of $\delta$ the second term of
equation (29) is negative and dominant. In sum, it seems that the
physically relevant region is where the exponent $|\delta| < 1/2$.

These approximations justify dropping terms of order $Z_0$ or
${Z_0}^\frac{1}{1+2\delta}$, and then the derivative of H is given by,

\begin{eqnarray}
\frac{\partial H}{\partial \delta}& = 0 = & 
\Bigl(49\delta^2 - 140\delta + 100\Bigr)
\Bigl(8\delta^3 + 12\delta^2 + 6\delta + 1\Bigr) \times
\nonumber\\
&&\biggl[-8\delta^4 - 4\delta^3 + 12\delta^2 - 194\delta + 32\biggr] +
\nonumber\\
&&2\Bigl(2\delta^2 - 2\delta + 11\Bigr)
\Bigl(49\delta^2 - 140\delta + 100\Bigr) \times
\nonumber\\
&&\Bigl(-4\delta^2 - 8\delta + 5\Bigr)
\Biggl(\frac{Z_0}{2}\Biggr)^{\frac{-2\delta}{1+2\delta}} +
\nonumber\\
&&32\delta \Bigl(1 - \delta\Bigr)
\Bigl(2\delta^2 - 2\delta + 11\Bigr)^2 \times
\nonumber\\
&&\Bigl(-7\delta + 10\Bigr)
\Biggl(\frac{Z_0}{2}\Biggr)^{\frac{2\delta}{1+2\delta}}.
\end{eqnarray}

This equation can be solved as a quadratic in the variable
$({Z_0}/2)^\frac{2\delta}{1+2\delta}$ as a function of $\delta$. The
value of $Z_0$ itself is then obtained for each value of $\delta$. We
find that equation (30) yields no positive, real values of $Z_0$ for
positive values of $\delta$ less than about 1/3. With the constraints
cited above this means that there is no significant range of positive
values of $\delta$ that give physical solutions. That is, solutions
with wind velocity increasing with radius do not satisfy the least
dissipation constraint. On the other hand, there are solutions to
equation (30) for negative values of $\delta$. The value of $Z_o$
increases monotonically (and $r_1/r_0$ decreases monotonically) as
$\delta$ goes from 0 to increasing negative values. If we make the
reasonable requirement that the radial extent of the outflow with $u <
v_e$, that is ${r_1}/{r_0}$, be greater than a few, then $|{\delta}|$
must be less than about 0.1. As the magnitude of $|{\delta}|$ is
decreased below 0.1, $r_1/r_0$ rises very rapidly, so the wind is very
large for such small values of $\delta$.

In sum, the Least Dissipation Theorem, as applied to the shock heating
function, has constrained the family of wind solutions. Specifically,
the parameter $\delta$ is limited to a small range for optimal
winds. Because the physically relevant values of $\delta$ have a very
small magnitude all the members of this one-parameter family are
nearly constant velocity winds. This result is not completely
rigorous, though physically reasonable. It is worth emphasizing that
because self-regulation is a global process, global constraints
determine the wind structure.

\subsubsection{Constraints from the Bernoulli Equation}

As an integral of the momentum equation, the Bernoulli equation
provides another global constraint on wind structure. For the
accelerating winds with wave pressure described equation (17), the
corresponding Bernoulli equation is,

\begin{equation}
\frac{{U^2}(r)}{2} + \frac{{c^2}(r)}{{\gamma'}-1} +
{\sigma^2}(r) - \frac{{v_e^2}}{2} = C_{\delta},
\end{equation}

\noindent where, as in equation (3), $v_e$ is the escape velocity,
$\gamma'$ is the global effective ratio of specific heats, and the
integration constant $C_{\delta}$ is a function of the wind
acceleration exponent $\delta$ only.

In the case of an accelerating wind we have expressions for all of
the terms on the left-hand-side of this equation, except the sound
speed term. Using equation (9) or (19) above, we find,

\begin{equation}
\frac{c^2}{\gamma'} =
\Biggl(\frac{GM-A}{(2+\delta)r}\Biggr)
\Biggl[{1} - {\delta}
\frac{\Bigl(1+{\frac{2B}{U_0^2}}\Bigr){U_0^2}{r_0}}{GM-A}
\biggl(\frac{r}{r_0}\biggr)^{1+2\delta}\Biggl].
\end{equation}

Since for $|{\delta}|$ small, we have
$P \propto \rho^{\frac{3+\delta}{2+\delta}}$, and then,
\begin{equation}
\gamma' = \frac{3+\delta}{2+\delta}.
\end{equation}

The value of the constant $C_{\delta}$ can be determined by evaluating
the left-hand-side of the Bernoulli equation at the inner edge. In
Bowen's numerical models of dust-free winds we find that,

\begin{equation}
\frac{c_0^2}{\gamma'-1} \simeq 2{c_0^2} <<
\frac{v_{e0}^2}{2}.
\end{equation}

\noindent The models also indicate that $\sigma_0 \le c_0$, and that
$U_0$ is much less than either of these. Thus, the terms on the
left-hand-side of equation (31) nearly cancel, and $|C_{\delta}|$ has
a relatively small value.

Next, we compare an accelerating wind, with $|{\delta}|$ small, but
nonzero, to the constant velocity wind having the same outflow
velocity at the inner wind radius. We also assume that the inner
radius and stellar mass are the same for both winds. Taking the
difference between the Bernoulli equations for the two winds we
obtain,

\begin{eqnarray}
\frac{1}{2} \bigl(U^2 - {U_0^2}\bigr) -
\delta\Biggl(\frac{GM-A}{r_0}\Biggr) \frac{r_0}{r} -&&
\nonumber\\
\biggl(\frac{\delta}{2+\delta}\biggr)
\Biggl(1 + \frac{2B}{U_0^2}\Biggr) {U_0^2}
\Biggl(\frac{r}{r_0}\Biggr)^{2\delta} + &&
\nonumber\\
B\Biggl(\frac{r}{r_0}\Biggr)^{2\delta} - B
&  = & C_{\delta} - C_{\delta=0}.
\end{eqnarray}

To determine the difference between constants on the right-hand-side
of this equation we evaluate it at $r = r_0$. We substitute the
result back into equation (35), and solve this equation for the
fractional difference in the mean flow velocity, as a function of r,
\begin{eqnarray}
\Biggl(\frac{U^2 - {U_0^2}}{{U_0^2}}\Biggr)& = &
2\delta \Biggl(\frac{GM-A}{{r_0}{U_0^2}}\Biggr)
\Biggl(\frac{r}{r_0} - 1\Biggr) +
\nonumber\\
&&\biggl(\frac{2\delta}{2+\delta}\biggr)
\Biggl(1 + \frac{2B}{U_0^2}\Biggr)
\Biggl[\biggl(\frac{r}{r_0}\biggr)^{2\delta} - 1\Biggr]
\nonumber\\
&& -  \frac{2B}{U_0^2}
\Biggl[\biggl(\frac{r}{r_0}\biggr)^{2\delta} - 1\Biggr].
\end{eqnarray}

Because $U_0^2$ is generally much less than the escape velocity
throughout the wind, except possibly the outermost parts, and because
the magnitude of $|B|$ is comparable to $U_0^2$, the first term on the
right-hand-side of this equation will generally dominate the others.
(Note that in the limit of small $|\delta|$, the last term is also
proportional to $\delta$, so it does not exceed the first term.)
However, the sign of the first term is such that it would imply a
falling U for a positive $\delta$, and a rising U for negative
$\delta$, which contradicts the definition $U \propto
r^{\delta}$. The only alternative is $\delta = 0$.

As before, this conclusion is not mathematically rigorous, because it
is not always true that the first term on the right-hand-side of the
above equation must dominate. For example, the parameter B goes to
positive infinity as $\delta$ approaches a value of 10/7. However,
such cases are exceptional, and unphysical. In sum, we generally
expect that {\it thermal winds driven by nonlinear acoustic waves have
constant outflow velocities.}

\subsection{Wave Pressure and the Outer Wind}

One remaining aspect of the dust-free winds to consider is what
happens in the outermost regions? For brevity, we will confine our
discussion to the constant velocity wind. The escape velocity $v_e$,
the sound speed c (eq.(32)), and the velocity dispersion associated
with the waves, $\sigma$ (eq. (18)) all decrease with as
$r^{\frac{-1}{2}}$ in this case. In the outer wind the constant flow
velocity will eventually equal and exceed each of these velocities in
turn.

The effective $\sigma$ is generally somewhat less than the sound speed
c, so we might expect a transition from the constant velocity wind to
a nearly adiabatic outflow before the material escapes.
Specifically, for the constant velocity wind with $\gamma' = 3/2$, the
parameter A = 2GM/5 (eq. (21)), and equations (18) and (32) yield,
\begin{displaymath}
\sigma_0 \simeq 0.45v_{e0}
\end{displaymath}
\begin{equation}
c_0 \simeq 0.47v_{e0}.
\end{equation}
Thus, we see that when $U_0$ equals $\sigma(r)$, it very nearly equals
c. However, these velocities are less than half the escape velocity,
and there is no reason to expect an abrupt increase of c(r). That is,
we do not expect a transition to supersonic flow through a Parker
critical point (where $U_0 = c = v_e$). Physically, as $U_0$ becomes
comparable to $\sigma$ and c, and all about equal to ${v_e}/2$, we
might expect successive pulsational waves to propel material above the local
escape velocity, and thus free of the star. It appears that the
outermost wind must be nonsteady.

The result that the sound speed is about half the local escape
velocity is consistent with the numerical models. However, the
velocity amplitude of the pulsations is much less than that. As
discussed above, this likely due in part to the limited spatial
resolution of the models. Moreover, the wave pressure used here could
well represent the aggregate effect of several pulsational waves. (In
fact, it is possible for one shock to overtake and merge with another,
though this process is not well-resolved in the numerical models.)

We can illustrate this last point with a more detailed comparison to
the numerical model. We consider what wave pressure boosts are
sufficient to push a gas element through a radius change of order
unity, ${\delta}r/r = 1$. The wave pressure drives the constant mean
flow, and the time for the gas element to flow this distance is
${\delta}t = {\delta}r/U$. The number of pulsational shocks passing
through the gas element in this time is about, $N = {\delta}t/T$,
where T is the pulsation period. We identify the net wave pressure
with the sum of the pressures in each shock to get, ${\sigma}^2 =
N{\Delta}v^2$. According to equation (37) the model predicts, $\sigma
\simeq 0.45v_e$, and so it also predicts,
\begin{equation}
{\Delta}v \simeq \frac{0.45}{\sqrt{N}} v_{e0}.
\end{equation}

As a specific example we take $r = {\delta}r = 10^{14} cm$, and $U =
0.6\ km/s$. This is a radius where the numerical model shows that
strong shocks present at smaller radii have settled into the steady
wind. There are still significant velocity fluctuations well beyond
this radius, but the adopted mean outflow velocity is a fair average
for the whole wind region. With these values and a pulsation period of
about one year, we compute the following values for the quantities
defined in the previous paragraph,
\begin{displaymath}
{\delta}t = 1.67 \times 10^9s,\ N \simeq 54,\ v_e = 18.\ km/s,\ 
\end{displaymath}
\begin{equation}
and\ 
{\Delta}v \simeq 1.1\ km/s.
\end{equation}
We set $\Delta v$ equal to this latter value at the adopted radius,
use the scaling of the constant velocity outflow model ($\sigma
\propto r^{-1/2}$), and add the mean outflow U to derive the lower
dashed curve in Figure 2. This curve is generally comparable to the
velocity jumps of the shocks in the inner wind, where these jumps are
fairly well resolved. In fact, the velocity jumps appear to be
somewhat larger than the curve in the inner wind, but this is
reasonable since some of the shock energy will go into thermal heating
as well as doing work against the star's gravity.

In the present example the work per mass of gas moved is
$-GM/(2r)$. The corresponding shock heating per gas mass, integrated
along the path, is ${\Gamma}{\delta}t = 0.11\ (GM/r)$. The integrated
work done by the wave pressure is about half what is needed. This
confirms the impression given by the figure that there is not too much
wave energy available for heating. In sum, in terms of the wind
driving energetics, wave pressure and thermal pressure play comparable
roles. A fraction of the thermal energy is replaced by shock
dissipation, as required for an effective adiabatic index of 3/2.

\section{Self-Regulation in Radiative and Dust Driven Winds}

If dust grains, atoms or molecules are able to absorb a fraction of
the substantial Mira radiative output, then radiation pressure can
play a substantial role in driving or accelerating the wind.  In the
case of typical Mira variables, the models of \citet{bow88}, and
\citet{bow91}, show that models with significant radiation pressure on
grains have higher mass loss rates and higher wind velocities. The
wind regions are also much smaller in these models because the wind
velocity surpasses the escape velocity at a smaller radii.

The pulsational hydrodynamics is essentially frozen out in a dusty
wind. There is, however, another regulatory process that limits the
wind velocity and keeps the flow from over-refrigerating. In fact, the
numerical models show that the gas temperature tracks the radiative
equilibrium temperature quite closely, which implies a coupling of
photons, grains and gas atoms. Atom-grain collisions are the mechanism
of both momentum and energy transfers (\citealt{bow88}). The gas flow
is driven by the momentum transfer from each grain to individual
atoms, and these are inevitably accompanied by energy transfers. The
interaction will also decelerate the grain, and will also result in
grain ablation if the relative velocity is too great. Ablation will
reduce the grain cross section, and the radiative driving. Slow wind
speeds allow more time for grain growth in dense regions, leading to
more radiative driving.

In particular, once the grains relax to a fixed size distribution, the
fraction of the photons they absorb, and the momentum input from them,
decrease as $1/r^2$. In this case, two terms dominate the
right-hand-side of a momentum equation, like equation (2), the
gravitational term and the radiative driving source term. These terms
scale the same way with radius, allowing a constant velocity outflow
solution. Other solutions to the momentum equation are
possible. However, because of the regulatory feedbacks, we expect that
the process of grain growth towards an equilibrium size distribution
should naturally be accompanied by wind acceleration to a constant
terminal velocity.

\section{Conclusions}

In summary, the numerical models indicate that the winds that develop
in the extended atmosphere of long-period variables have a simple
power law structure. The isentropic Parker wind solution that matches
this structure has a barotropic index of $\gamma = 3/2$. In a real gas
heating is required to maintain this index, but the Parker solution
does not provide information about the heating source. In the case of
non-dusty Miras, where radiation pressure is unimportant, we believe
that the power law structure is generated and supported by shock waves
which travel through the wind. Since shocks dissipate energy and
generate entropy, these winds have significant heat inputs and entropy
gradients. In section 2 we presented phenomenological equations
including the three relevant terms: wave pressure, wave heating and
large-scale entropy gradients. We then studied a family of analytic
thermal wind solutions to these equations with power law velocity
profiles. These wind solutions generalize the classical isentropic
Parker solutions to cases in which the gas is only locally adiabatic.
The simplest member of this family has a constant outflow velocity,
and matches numerical models of non-dusty Mira winds quite well.

Section 2 concluded with a discussion of why the constant outflow
velocity solution may be preferred in nondusty Miras. Both the Least
Dissipation Theorem of linearized, nonequilibrium thermodynamics, and
the Bernoulli equation with reasonable physical constraints indicate
that this is a preferred solution among the family of power law
winds. This result provides reassurance of the basic correctness of
Bowen's numerical models, and the validity of mass loss estimates
predicted by them.

In Section 3 we considered dusty winds in long-period variables. In
this case the winds are driven by radiation pressure on dust grains,
and the flow ``freezes out'' in the sense that thermal, wave-driven or
turbulent pressures are not dynamically important. This allows the
wind velocity to attain supersonic speeds without passing throught the
critical point of classical thermal wind theory. Nonetheless, Bowen's
numerical models show that there are significant couplings between the
radiation field, an equilibrium distribution of dust grains, and the
atomic gas.  These couplings allow a regulated, constant velocity
outflow to form and be maintained.

We believe that the model of dust-free Mira winds may provide a
paradigm, and that gas dynamic solutions with acoustic wave pressure
may have much more general application in astrophysics. Specifically,
such solutions will be relevant whenever there is substantial
supersonic or magnetosonic (velocity) noise on small scales as
compared to large scale velocities and thermal pressure gradients.
Laboratory examples are provided by the reverberating shock cavities
described in \citet{hol95} and \citet{wei96}).  Other possible
astrophysical examples include: 1) accretion disks with strong
acoustic-convective turbulence, 2) galactic winds or outflows driven
by continuing supernova shocks, and 3) the hot gas halos of galaxy
clusters, containing galaxies moving at transonic or supersonic
velocities, and into which there is continuing infall. The defining
property is that the entropy or the function K have a nonzero
gradient.

\section*{Acknowledgments}
We are grateful to S. Owocki for helpful conversations.

\appendix

\section{Appendix A: Density and Velocity Averages}

In this appendix we give a brief, qualitative discussion of how the
density and velocity fluctuations driven by pulsational shocks
propagating through an LPV wind can be divided into a mean and a
fluctuating part that averages to nearly zero over a pulsational
period. We also discuss the derivation of the Reynolds equations for
this flow. To begin, consider the trajectories of a couple of adjacent
gas elements relative to the local mean flow.  We choose the gas
elements that are sufficiently close together that the local mean
outflow can be approximated as constant velocity, even if it is not
globally constant.

Next, we divide the pulsational cycle into two parts.  The first part
begins when the gas element is hit by a shock.  This impact boosts the
fiducial element outward, and after a brief delay does the same to the
neighboring element at slightly larger radius.  They are pushed closer
together in the shock compression, so both $\delta\rho$ and $\delta$u
$> 0$, i.e., greater than the values of the mean flow (see e.g.,
Fig. 2 of \citealt{bow88}.

The fiducial gas element is closer to the star than its neighbor, and
so, has a shorter free-fall time.  As a result of this and the fact
that it received its outward velocity burst first, it will decelerate
ahead of its neighbor.  This ballistic description is equivalent to
the statement that the shock is followed by a rarefaction wave.  We
define the beginning of the second part of the cycle as beginning at
the moment when the velocity is reduced to the mean flow value, and
continuing until the next shock impact. The density is reduced to the
mean value at about the same time. Thus, during this second interval
we have, $\delta\rho$, and $\delta$u $< 0$.

Because the pulsational shocks are not very strong in the wind region,
the magnitudes of $\delta\rho$, and $\delta$u are not large, and the
cycle is nearly symmetric in the sense that the two intervals are
about equal to half a cycle.  Thus, the net density and velocity
fluctuations over one cycle are small, i.e., of higher order than
$\delta$r/r, the fractional radius change in that time. This justifies
the Reynolds procedure of writing each gas quantity as a sum of a mean
and a fluctuating part. E.g., for the velocity $u = U + \sigma$, and
for the other fluid variables, $\rho = <\rho> +\ {\rho}'$, $P = <P> +\
P'$, $\epsilon = <\epsilon> +\ {\epsilon}'$, etc., where
the primes denote the fluctuating parts.

The Reynolds equations are then derived by substituting these
variables into the hydrodynamical equations and averaging over a time
greater than the pulsational period. It is assumed that the
mean flow still satisfies the hydrodynamical equations. Then,
subtracting the mean flow equation from the corresponding equation for
mean plus fluctuating parts yields an equation for the fluctuating
part. (See Section 1.3 of \citealt{mcc90}, which the following
discussion parallels.)

For example, the spherically symmetric steady state continuity
equation is,
\begin {equation}
\frac{\partial}{\partial r}
\biggl(r^2 <{\rho}u> \biggr) = 0.
\end {equation}
The mean flow continuity equation is,
\begin {equation}
\frac{\partial}{\partial r}
\biggl(r^2 <{\rho}>U \biggr) = 0
\end {equation}
(see eq. (1)), and the fluctuation continuity equation is,
\begin {equation}
\frac{\partial}{\partial r}
\biggl(r^2 <{\rho}'>U +
r^2 <<{\rho}>{\sigma}> +
r^2 <{\rho}'{\sigma}>\biggr)
= 0.
\end {equation}

On averaging this last equation we find,
\begin {equation}
<{\rho}'>U = 0,\ \ <<{\rho}>{\sigma}> = <{\rho}><{\sigma}> = 0,
\end {equation}
so we have,
\begin {equation}
<{\rho}'{\sigma}> = 0.
\end {equation}

The steady velocity equation is,
\begin {equation}
u\frac{\partial u}{\partial r} =
\frac{-1}{\rho} \frac{\partial P}{\partial r} -
 \frac{\partial <{\sigma}'_r>}{\partial r} -
\frac{GM}{r^2},
\end {equation}
where in contrast to equation (2) we have included a mean stress
gradient, or wave pressure, term (e.g., \citealt{lan59}). After we
subtract the mean flow version of this equation, we get the following
equation for the fluctuating velocity (before averaging),
\begin {eqnarray}
u'\frac{\partial U}{\partial r} +
&U\frac{\partial u'}{\partial r} +
\frac{\partial}{\partial r} \biggl({\sigma}_r^2 - <{\sigma}_r^2>\biggr) 
 = \frac{-1}{<\rho>} \frac{\partial P'}{\partial r} +\\
&\frac{1}{<\rho>} \frac{{\rho}'}{<\rho>} 
   \frac{\partial <P>}{\partial r}
+ higher\ order\ terms.
\end {eqnarray}
As before, all of the first order terms will vanish on averaging,
leaving only thermal pressure and density terms at second order, which
we assume are negligible.

Equation (4) of Section 2.2.2 is the total energy
equation. Writing each variable in terms of its mean and fluctuating
parts, and averaging with the assumption that odd order moments
average to zero, yields equation (5). With the additional
approximations described in that section, this equation reduces to
equation (6).

The various approximations provide a simple closure of the Reynolds
moment equations in this application, but they do so via the
assumption that dissipation in the wind formation region has taken the
system to a relaxed state characterized by small
fluctuations. Therefore, the equations do not provide a valid
description of the unrelaxed dynamics at the onset of pulsations, nor
of the strong shock zone shown by numerical models to exist at smaller
radii (see Fig. 1-3).

\section{ Appendix B: Spherical Shocks Propagating Down a
Power Law Density Gradient}

As described in the text, the numerical models show that the extended
atmospheres and winds of LPVs relax to power law density and pressure
profiles, and in fact, the main result of this paper is on how these
profiles can be accounted for with the aid of shock pumping and mean
entropy gradients. These conclusions are founded on the assumption
that shocks propagate down the relaxed power law pressure gradient
without significantly accelerating (or decelerating). In this appendix
we briefly review some classical results that demonstrate the
existence of profiles for constant shock propagation. These results do
not seem to be well-known in the astrophysical literature, though they
are summarized in the text of \citet{whi74}.

The first result concerns the outward propagation of a spherically
symmetric shock wave in a constant density medium. The wave is not
necessarily a strong blast as assumed in the well-known Sedov-Taylor
solution. An approximate similarity solution to this more general case
was presented by \citet{gud42}, also see Section 6.16 of
\citet{whi74}. Specifically, in the spherical adiabatic case the
post-shock velocity is found to vary as,

\begin{equation}
p \propto r^{-0.905},
\qquad
u \propto r^{-0.453}.
\end{equation}

Density gradients affect shocks in a manner similar to geometric
convergence or divergence, and a related result of \citet{sak60},
described in Section 8.2 of \citet{whi74} is useful. That is, the
velocity of a planar adiabatic shock on a density gradient, $\rho_o
\propto x^{\alpha}$, varies as,

\begin{equation}
u \propto  x^{-\lambda},
\qquad
with
\quad
\lambda = {\alpha}{\beta},
\quad
\beta \simeq 0.236,
\end{equation}

\noindent and thus, $\lambda \simeq -0.472$ for $\alpha =
-2$. Therefore, combining these two effects, we expect the variation
of the spherical shock velocity on the density gradient $\rho_o
\propto r^{-2}$, to be roughly,

\begin{equation}
u \propto r^{0.472}r^{-0.453}
\simeq r^{0.02}.
\end{equation}

\noindent That is, the shock velocity is not exactly constant, but it
is essentially so within the accuracy of the approximations.

\label{lastpage}
\end {document}